\documentclass[preprint,12pt]{elsarticle}

\usepackage{amssymb}

\usepackage{lineno}

\journal{Nuclear Physics A}

\begin{document}

\begin{frontmatter}



\title{Azimuthal Anisotropy Measurements by STAR}


\author{Li Yi for STAR collaboration}

\address{Department of Physics, Purdue University, West Lafayette, IN, 47906, USA}

\begin{abstract}
The recent study of centrality and transverse momentum ($p_{T}$) dependence of inclusive charged hardron elliptic anisotropy ($v_{2}$) at midrapidity ($|\eta|<1.0$) in Au+Au collision at $\sqrt{s_{NN}} =$ 7.7, 11.5, 19.6, 27, and 39 GeV in STAR Beam Energy Scan program is presented. We show that the observed increase of inclusive $v_{2}$ is mainly due to the average $p_{T}$ increase with energy. In Au+Au 200 GeV collisions, the triangular anisotropy ($v_{3}$) measurements highly depend on measurment methods; $v_{3}$ is strongly dependent on $\Delta\eta$. The difference between two- and four-particle cumulants $v_{2}\lbrace{2}\rbrace$ and $v_{2}\lbrace{4}\rbrace$ for Au+Au and Cu+Cu collision at $\sqrt{s_{NN}} = $ 62.4 and 200 GeV is used to explore flow fluctuations. Furthermore, by exploiting the symmetry of average flow in pseudorapidity $\eta$ about mid-rapidty, the $\Delta\eta$-dependent and independent components are seperated using $v_{2}\lbrace2\rbrace$ and $v_{2}\lbrace{4}\rbrace$.

\end{abstract}


\end{frontmatter}


\section{Introduction}
\label{sec:intro}

Azimuthal anisotropies of particle distribution in high energy heavy-ion collisions provide information about the collision dynamics \cite{Ollitrault}. These anisotropies, $v_{n}$, are sensitive to the initial state geometry, the equation of state, and transport coefficients of the medium \cite{Shen}. 

In this paper, the dependence of the azimuthal anisotropies on collision energy, centrality, $p_{T}$ and $\Delta\eta$ are studied. The effects  of different measurement methods are also discussed. The nonflow and flow fluctuation effects in anisotropy measurements are explored. An attempt to isolate nonflow from flow is made by using the symmetry about midrapidity in Au+Au collisions.

\section{Elliptic Anisotropy Beam Energy Dependence}
\label{sec:v2}

\begin{figure}[htb]
\centering
\includegraphics[width=0.8\textwidth]{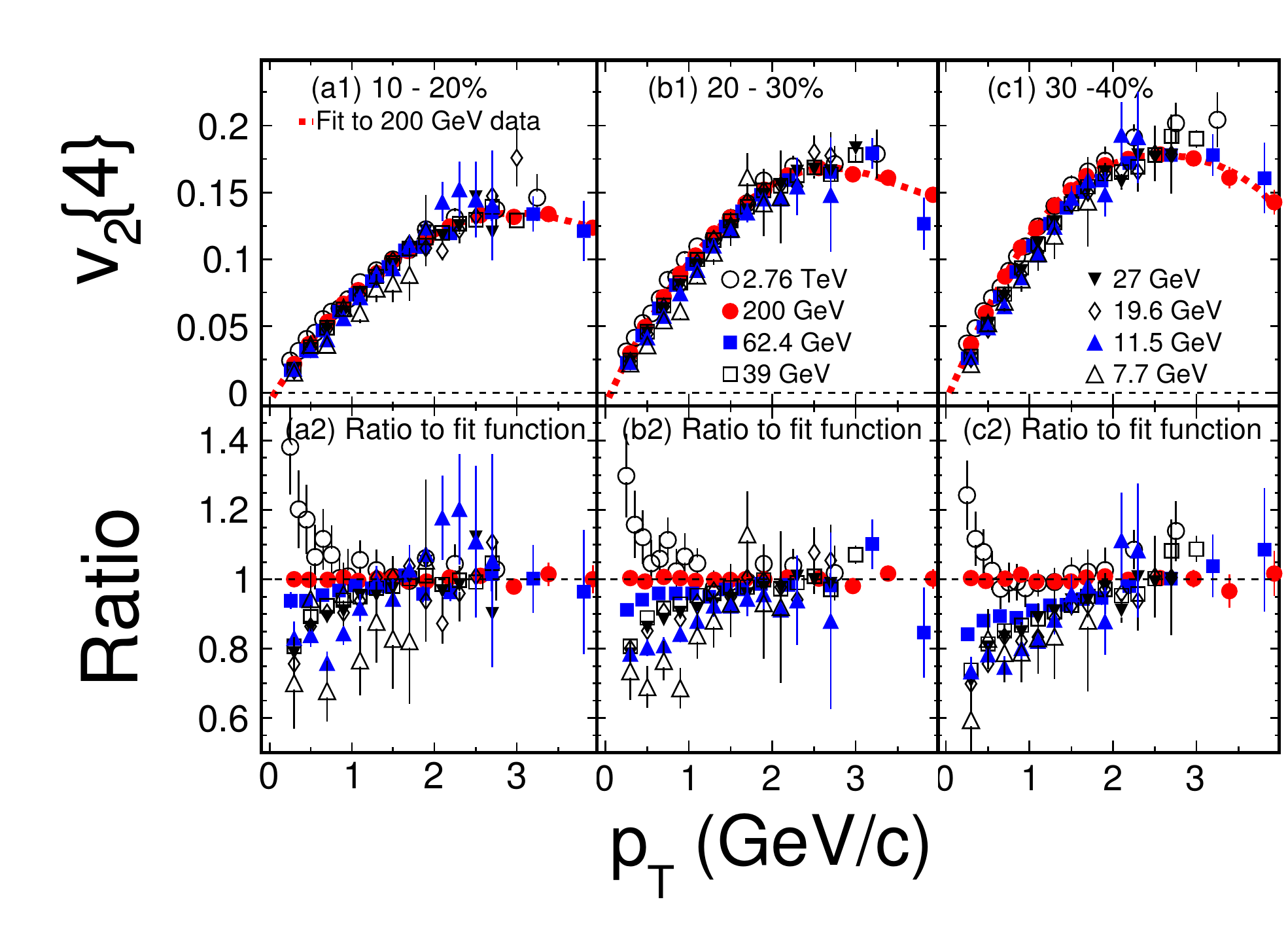}
\caption{The top panels are $v_{2}\lbrace4\rbrace$ as a function of $p_{T}$ in $\sqrt{s_{NN}}$ = 7.7 GeV to 200 GeV Au+Au collisions and 2.67 TeV  Pb+Pb collisions, for centralities: (a1) 10-20\%, (b1) 20-30\%, (c1) 30-40\%. The dashed red curves are empirical fits to the 200 GeV Au+Au data. The bottom panels show the ratio of $v_{2}\lbrace4\rbrace$ to the fit for all beam energies. }
\label{fig:BESv24pt}
\end{figure}

The increase of the $p_{T}$ integrated $v_{2}$ with beam energy from AGS \cite{E877:v2} to LHC \cite{ALICE:v24} could be due to an increase of $v_{2}(p_{T})$, the increase of the mean transverse momentum $\langle p_{T} \rangle$ ( or a change of particle composition). The RHIC Beam Energy Scan (BES) program enables a systematic study of the differential azimuthal anisotropy as a function of collsion energy (7.7 GeV, 11.5 GeV, 19.6 GeV, 27 GeV and 39 GeV). $v_{2}\lbrace4\rbrace$ is insensitive to nonflow correlations so that it is chosen for the comparison in Figure~\ref{fig:BESv24pt}. Its dependence on $p_{T}$ is presented for 10-20\% centrality in panel (a1), 20-30\% (b1), 30-40\% (c1) \cite{STAR:BES}. The symbols represent Au+Au STAR results from BES and ALICE Pb+Pb results at $\sqrt{s_{NN}}$ = 2.76 TeV. The red curves are fifth-order polynomial function fit to STAR 200 GeV $v_{2}\lbrace4\rbrace$. The ratios of $v_{2}\lbrace4\rbrace$ to the fit curves are plotted in the lower panels for all collision energies. At $p_{T}<$ 1 GeV/$c$, $v_{2}\lbrace4\rbrace$ increases as collision energy increases, but the variations are less than $\pm$30\%. At $p_{T}>$ 1 GeV/$c$, the difference between various collision energies is less than $\pm$10\%. This suggests that the increase in the integrated $v_{2}$ is largely due to the increase of produced particle mean $p_{T}$. In the viscous hydrodynamic simulations with a constant shear viscosity to entropy density ratio $\eta/s$ and zero net baryon density, the splitting between different energies increases with $p_{T}$ \cite{Shen}, which cannot reproduce the $v_{2}\lbrace4\rbrace(p_{T})$ results.  

\begin{figure}[htb]
\centering
\includegraphics[width=0.8\textwidth]{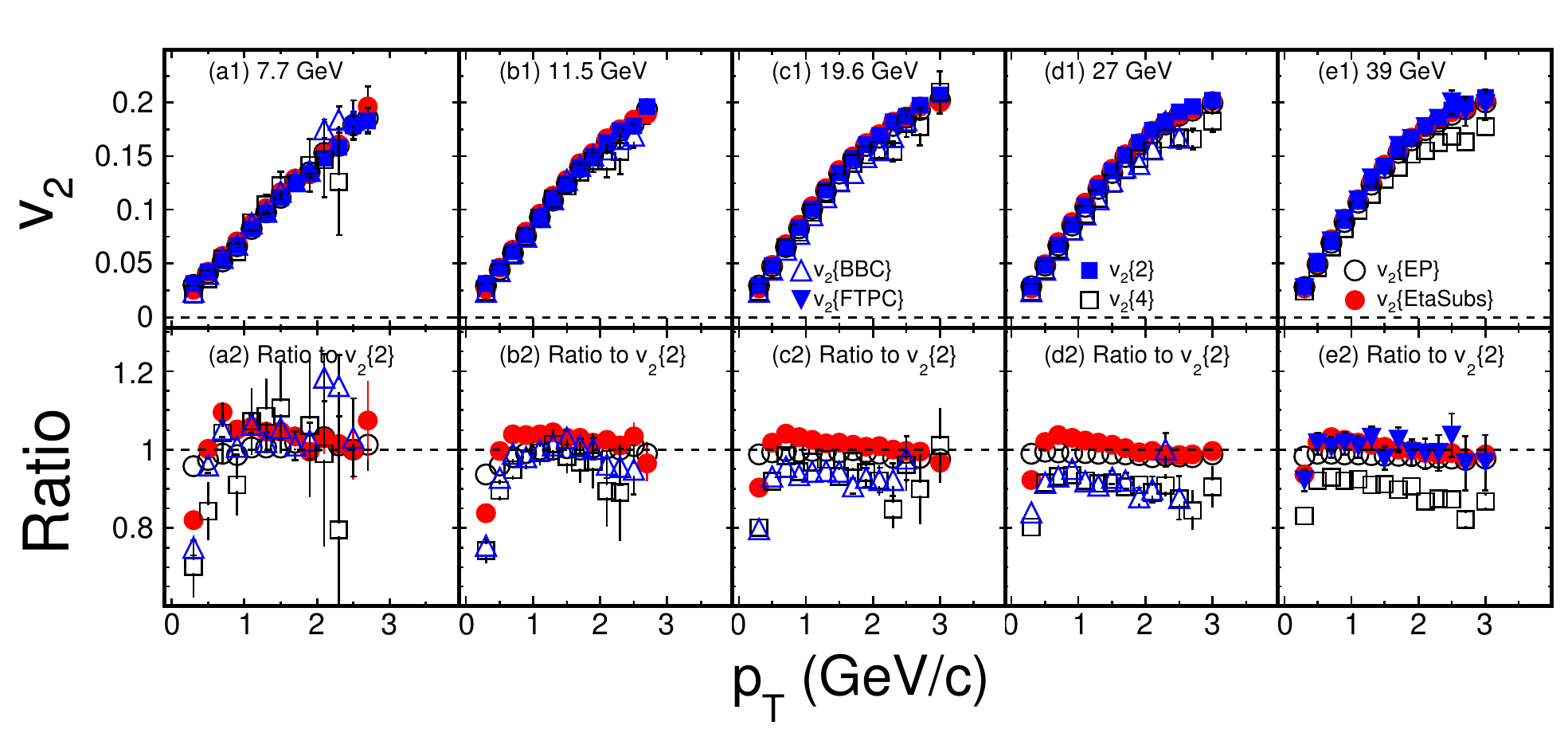}
\caption{The top panels are the $p_{T}$ dependences of $v_{2}$ in 20-30\% Au+Au collisions at $\sqrt{s_{NN}}$ = 7.7 GeV to 39 GeV from (a1) to (e1). The blue solid squares are $v_{2}\lbrace2\rbrace$; The black open squares are $v_{2}\lbrace4\rbrace$. The bottom panels show the ratio of $v_{2}$ from different methods to the two-particle cumulant $v_{2}\lbrace2\rbrace$. }
\label{fig:BESv22v24}
\end{figure}

In Figure~\ref{fig:BESv22v24}, the top panels show the method comparison of the inclusive charged hadrons $v_{2}$ in Au+Au collisions from $\sqrt{s_{NN}}$ = 7.7 GeV (a1) to 39 GeV (e1). The two-pariticle $v_{2}$ is contaminated by nonflow. Meanwhile, the flow fluctuation effects are different in two- and four-particle $v_{2}$. Hence, the differences between $v_{2}\lbrace2\rbrace$ and $v_{2}\lbrace4\rbrace$ in Figure~\ref{fig:BESv22v24} indicate effects of nonflow and flow fluctuations at various energies. The ratios of $v_{2}$ from all methods to $v_{2}\lbrace2\rbrace$ are presented in the bottom panels. The $v_{2}\lbrace4\rbrace/v_{2}\lbrace2\rbrace$ ratio, represented by the black open squares, deviates from unity as energy increases. This suggests smaller nonflow and/or flow flucutations at the lower beam energies.

\section{Triangular Anisotropy in $\sqrt{s_{NN}}$ = 200 GeV Au+Au Collisions}
\label{sec:v3}

\begin{figure}[htb]
\centering
\includegraphics[width=0.4\textwidth]{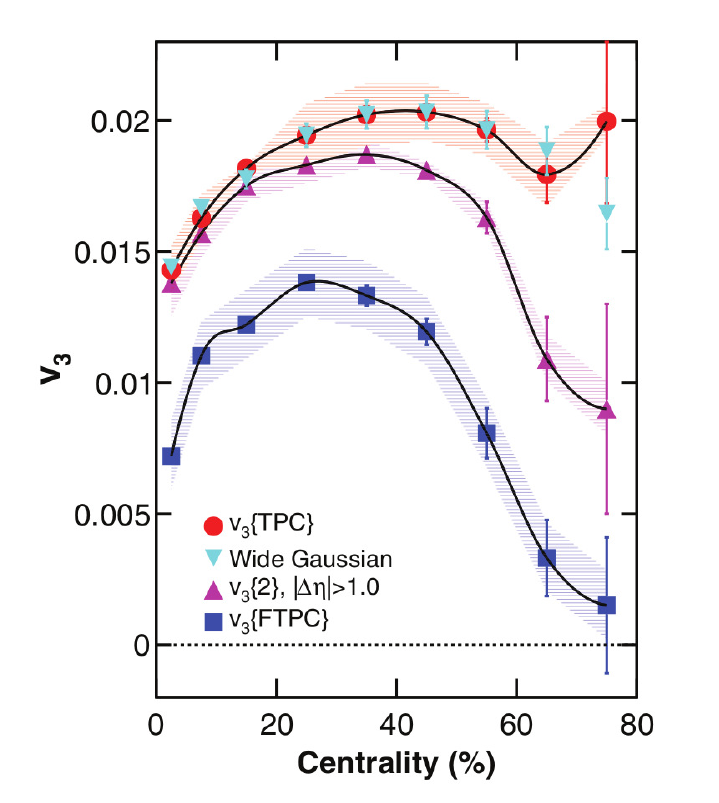}
\caption{The integrated triangular anisotropy $v_{3}$ as a function of centrality in $\sqrt{s_{NN}}$ = 200 GeV Au+Au collisions. The red dots represent TPC event-plane $v_{3}$. The blue solid squares are FTPC event-plane $v_{3}$. The cyan upside-down triangles are the fit results of a wide Gaussian to 2-particle $\Delta\eta$. The purple triangles are the 2-particle cumulant $v_{3}\lbrace2\rbrace$ with $|\Delta\eta|>$1.0. }
\label{fig:v3cent}
\end{figure}

The higher-order odd harmonic flow carries valuable information about lumpiness in the initial state of the colliding system \cite{Mishra}. The centrality dependence of integrated third harmonic $v_{3}$ of the charged particles with $|\eta|<1$ is presented in Figure~\ref{fig:v3cent} for Au+Au collisions at $\sqrt{s_{NN}}$ = 200 GeV \cite{STAR:v3}. The TPC event plane $v_{3}\lbrace\textrm{TPC}\rbrace$ (red dots) is about 30\% larger than the FTPC event plane $v_{3}\lbrace\textrm{FTPC}\rbrace$ (blue squares) for all centrality classes. Since $v_{3}\lbrace\textrm{TPC}\rbrace$ and $v_{3}\lbrace\textrm{FTPC}\rbrace$ are measured relative to the third harmonic event plane reconstructed in the TPC ($|\eta|<1.0$) and FTPC ($2.5<|\eta|<4.0$) subevents, respectively, their effective $\Delta\eta$ are different.

\begin{figure}[htb]
\centering
\includegraphics[width=0.6\textwidth]{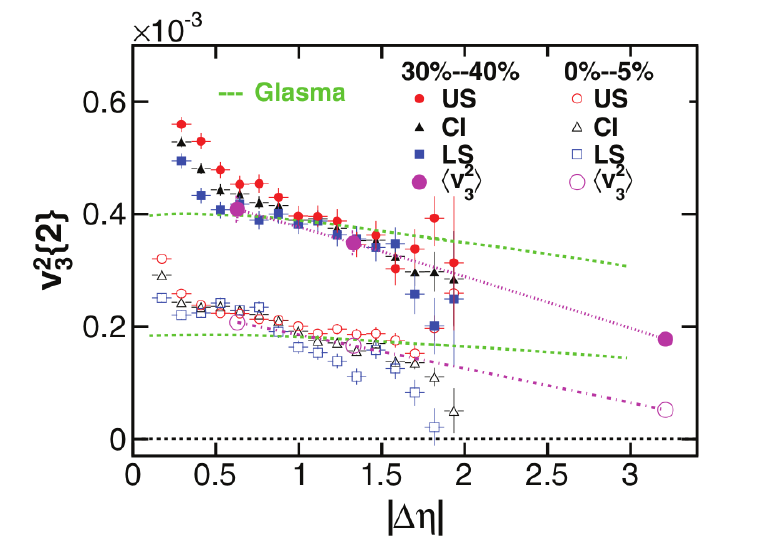}
\caption{The square of the two-particle triangular anisotropy as a function of pseudorapidity separation in $\sqrt{s_{NN}}$ = 200 GeV Au+Au collisions for 0-5\% (open symbols) and 30-40\% (solid symbols) centralities. The red symbols are for unlike sign, blue for like sign, and black for inclusive charges. Also shown, as the larger purple symbols, are the squares of the mean $v_{3}$ from the three measurement methods: $|\Delta\eta|$ = 0.63 for TPC event-plane $v_{3}$, 1.33 for two-particle cumulant with $\eta$-gap=1, 3.21 for FTPC event-plane $v_{3}$. }
\label{fig:v3deta}
\end{figure}

The two particle correlation $v_{3}^{2}\lbrace2\rbrace$ in the TPC as a function of $|\Delta\eta|$ is presented in Figure~\ref{fig:v3deta} for two centrality classes (open symbols for 0-5\%, and closed for 30-40\%) and three charge combinations (red for same sign charge, blue for opposite sign charges, and black for inclusive charge). For all the cases, there are strong $\Delta\eta$-dependence for $v_{3}$. It could be due to a dependence of the single particle $v_{3}$ on $\eta$. It could also be as a result of nonflow and/or flow fluctuation effects due to event-plane decorrelation over $\Delta\eta$ \cite{Xiao}.  Also shown on the plot are the results from the analysis in Fig.~\ref{fig:v3cent} as a function of the mean $|\Delta\eta|$. The point at $|\Delta\eta| = 0.63$ is from the subevent method using the TPC with $|\eta|<1$. The point at 3.21 is from that using the FTPC event plane. They generally agree with the gradual decrease of $v_{3}$ with $\Delta\eta$.

\section{Flow Fluctuations and Nonflow}
\label{sec:nonflow}

\begin{figure}[htb]
\centering
\begin{minipage}[b]{0.4\textwidth}
\includegraphics[width=\textwidth]{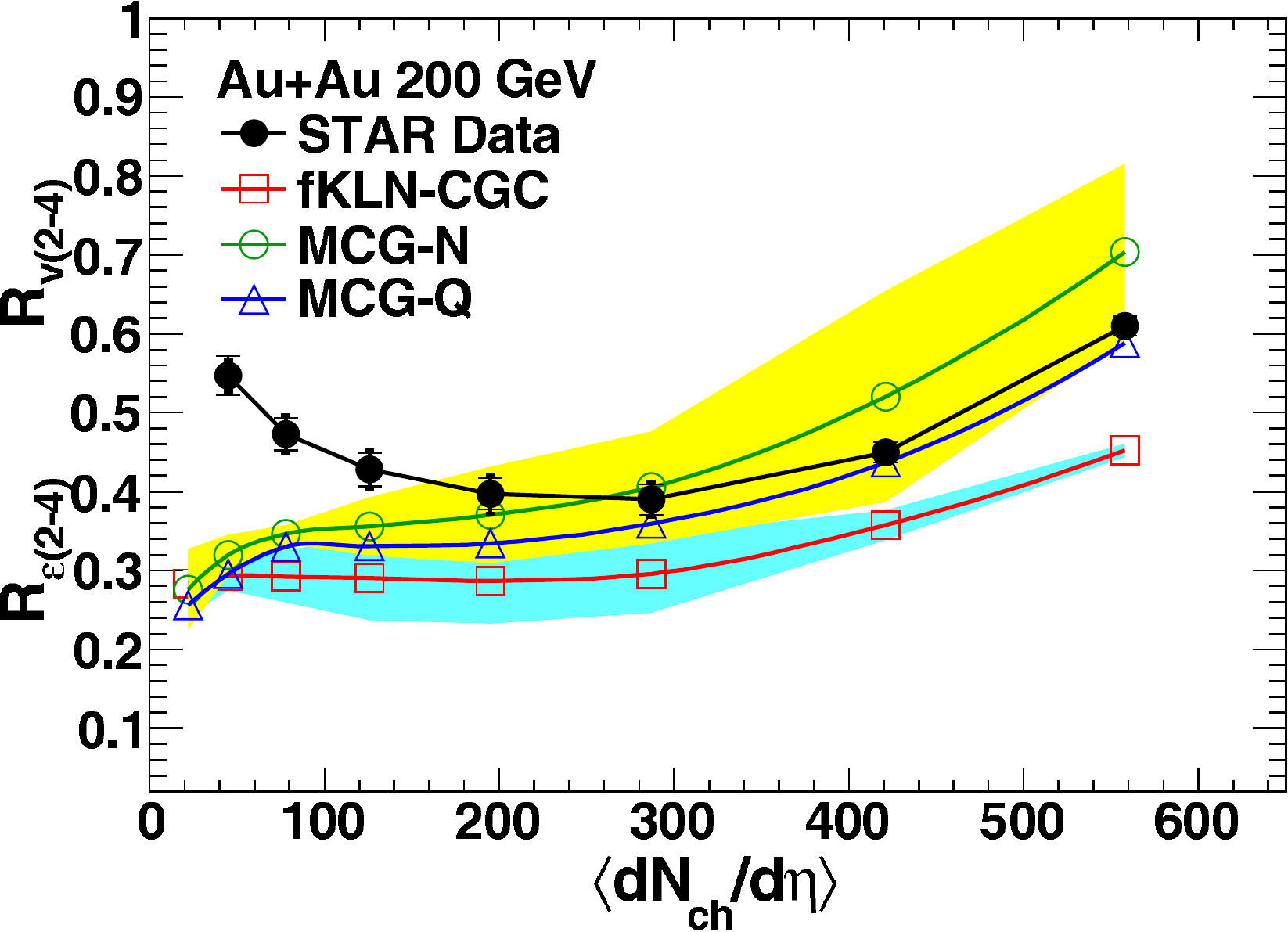}
\end{minipage}
\begin{minipage}[b]{0.4\textwidth}
\includegraphics[width=\textwidth]{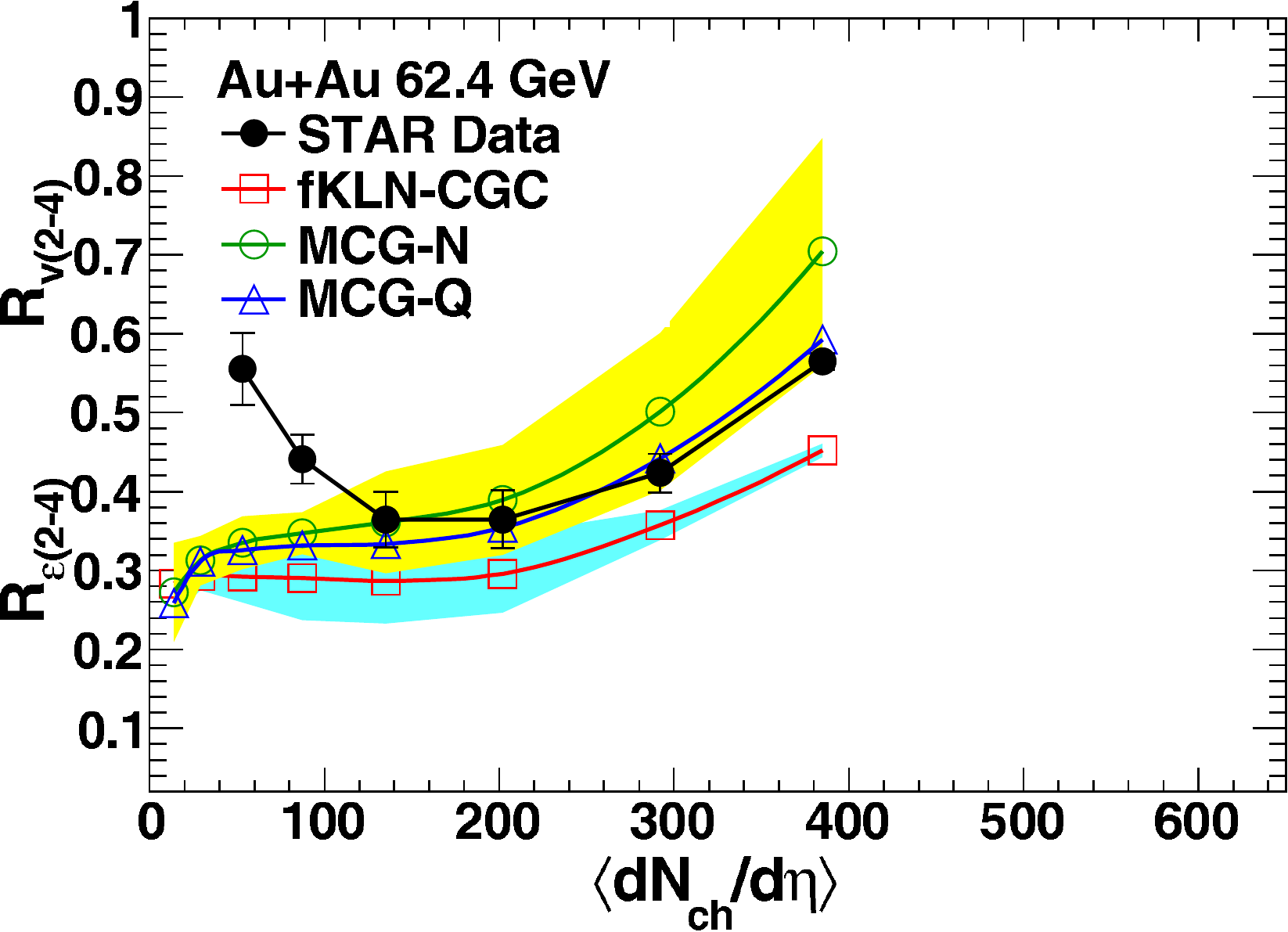}
\end{minipage}
\caption{The upper limit on $\sigma_{2}/\langle v_{2} \rangle$ in Au+Au collision at $\sqrt{s_{NN}}$ = 200 GeV (left) and 62.4 GeV (right) are shown as black dots. For comparison, $\sigma_{\epsilon}/\epsilon$ from three different models are also shown.}
\label{fig:upperlimit}
\end{figure}

Since $v_{2}\lbrace2\rbrace^{2} - v_{2}\lbrace4\rbrace^{2} \approx \delta_{2} + 2 \sigma_{2}^{2}$ ($\delta_{2}$ is nonflow, and $\sigma_{2}$ is flow fluctuation), the upper limit on the ratio $\sigma_{2}/v_{2}$ may be derivated by assuming $\delta_{2} \ge 0$. Figure~\ref{fig:upperlimit} shows $R_{v(2-4)} = \sqrt{\frac{v_{2}\lbrace2\rbrace^{2}-v_{2}\lbrace4\rbrace^{2}}{v_{2}\lbrace2\rbrace^{2}+v_{2}\lbrace4\rbrace^{2}}}$, the upper limit on the ratio $\sigma_{2}/\langle v_{2}\rangle$, in Au+Au collisions at $\sqrt{s_{NN}} = $ 200 GeV (left panel) and 62.4 GeV (right panel) \cite{STAR:ecc}. Three eccentricity models are shown on the plot as $R_{\epsilon(2-4)}$. The difference between $v_{2}\lbrace2\rbrace$ and $v_{2}\lbrace4\rbrace$ may have additional fluctuation sources other than eccentricity fluctuations. In periphral collisions, all three models are below the data points as expected. In central collisions, the ratio from MCG-N model seems to be the closest to the upper limit measurement. However, the measurement is an upper limit and there should be additional fluctuation sources besides eccentricity fluctuations, it is  premature to conclude which model is favored by the data. 

A recent study of $v_{2}\lbrace2\rbrace$ and $v_{2}\lbrace4\rbrace$ attempts to separate flow and nonflow in a data-driven way \cite{LY}. By exploiting the symmetry of average flow in $\eta$ about mid-rapidity in symmetric heavy ion collisions, the $\Delta\eta$-dependent and independent contributions may be separated \cite{Xu}. We associate the $\Delta\eta$-independent part with flow, and the $\Delta\eta$-dependent part with nonflow. This is because flow is an event-wise many-particle azimuthal correlation, reflecting properties on the single-particle level. On the other hand, nonflow is a few-particle azimuthal correlation, dependending on $\Delta\eta$ distance between particles.

By taking the difference between 2-particle cumulant $V_{n} \lbrace 2 \rbrace$ and the square root of 4-particle cumulant $V_{n} \lbrace 4 \rbrace$ at $(\eta_{\alpha}, \eta_{\beta})$ and $(\eta_{\alpha}, -\eta_{\beta})$, respectively, we have,
\begin{eqnarray}
\Delta V \lbrace 2 \rbrace & \equiv & V \lbrace 2 \rbrace (\eta_{\alpha},\eta_{\beta})- V \lbrace 2 \rbrace (\eta_{\alpha},-\eta_{\beta}) \nonumber \\
 & \equiv & V \lbrace 2 \rbrace (\Delta\eta_{1})- V \lbrace 2 \rbrace (\Delta\eta_{2}) = \Delta \sigma '  + \Delta \delta, \label{EqDV2} \\
\Delta V \lbrace 4 \rbrace & \equiv & V \lbrace 4 \rbrace (\eta_{\alpha},\eta_{\beta})- V \lbrace 4 \rbrace (\eta_{\alpha},-\eta_{\beta})  \nonumber \\
 & \equiv & V \lbrace 4 \rbrace (\Delta\eta_{1})- V \lbrace 4 \rbrace (\Delta\eta_{2}) \approx -\Delta \sigma ', \label{EqDV4}
\end{eqnarray}
where $
\Delta\eta_{1}  \equiv \eta_{\beta} - \eta_{\alpha}  , \ 
\Delta\eta_{2} \equiv -\eta_{\beta} - \eta_{\alpha} \label{EqDh} $ and $
\Delta\sigma'=\sigma'(\Delta\eta_{1})-\sigma'(\Delta\eta_{2}), \Delta\delta = \delta(\Delta\eta_{1})-\delta(\Delta\eta_{2})  \ 
$; $\sigma'$ denotes the $\Delta\eta$ dependent part of flow flucutation; $\delta$ is nonflow. The harmonic order $n$ is omitted from the equations for simiplicity. In symmetric heavy ion collisions, the $\Delta\eta$-independent terms in $V_{2} \lbrace 2 \rbrace$ and $V_{4}\lbrace 4 \rbrace$ cancel out. What remain in the differences are the $\Delta\eta$-dependent terms: flow fluctuation $\Delta \sigma'$ and nonflow $\Delta \delta$.  

\begin{figure}[htb]
\centering
\begin{minipage}[b]{0.4\textwidth}
\includegraphics[width=\textwidth]{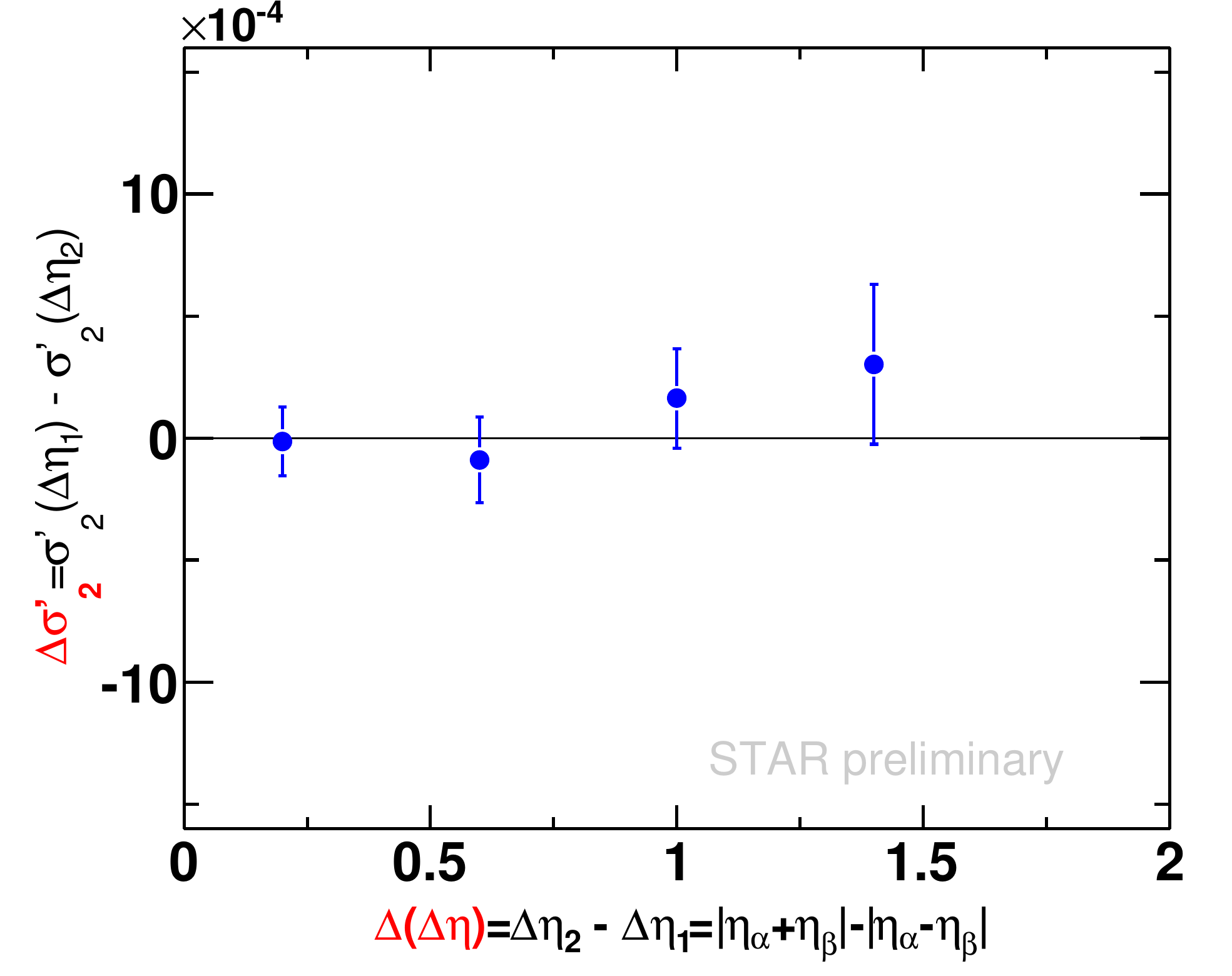}
\end{minipage}
\begin{minipage}[b]{0.4\textwidth}
\includegraphics[width=\textwidth]{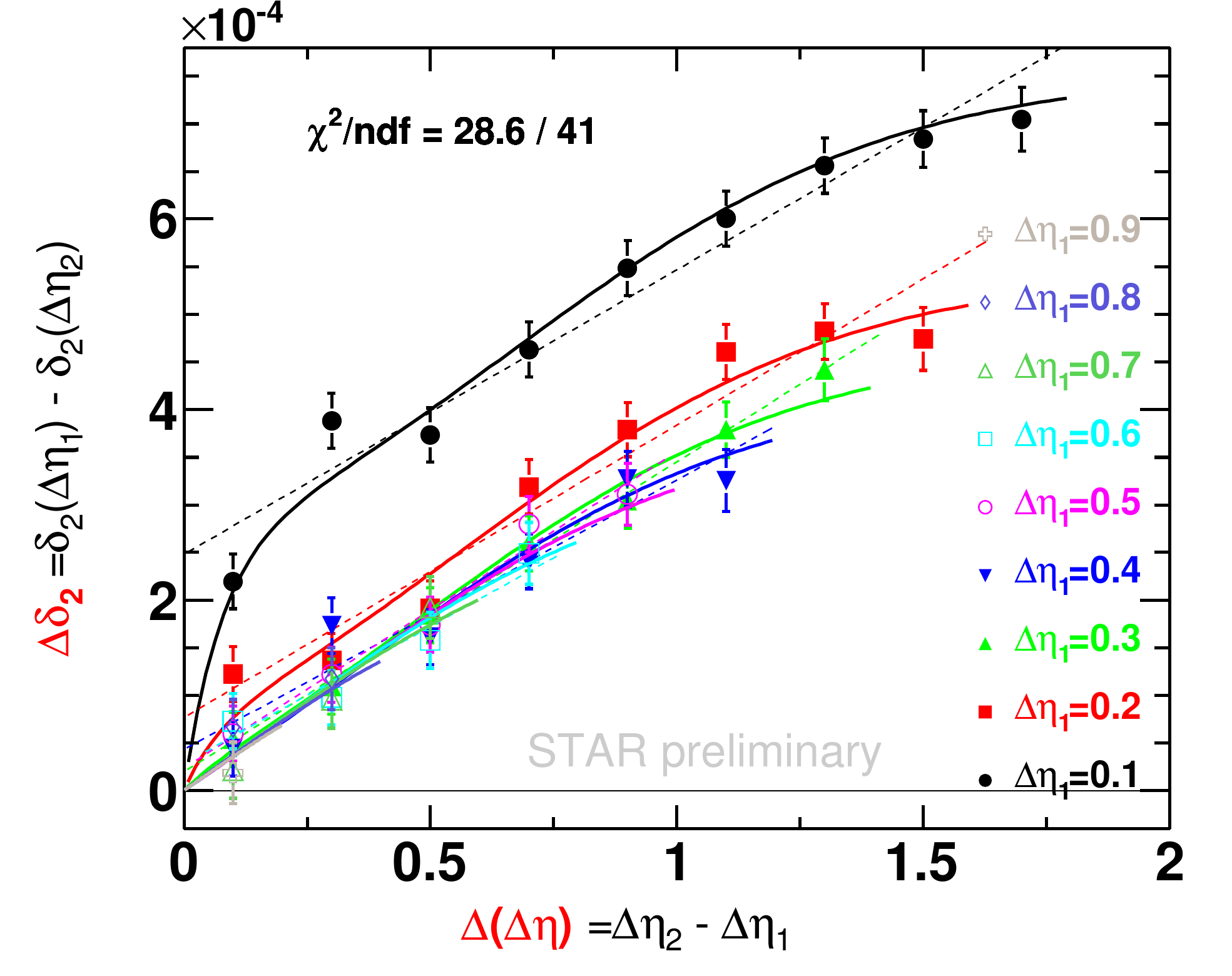}
\end{minipage}
\caption{Difference in $V_{2}\lbrace4\rbrace$ (left) and $V_{2}\lbrace2\rbrace$ (right) between ($\eta_{\alpha},\eta_{\beta}$) and ($\eta_{\alpha},-\eta_{\beta}$) as a function of pseudorapidity difference between these two pairs two particles pseudorapidity separation $\Delta\eta_{2} - \Delta\eta_{1} = |2 \eta_{\beta}|$ for 20-30\% at $\sqrt{s_{NN}}$ = 200 GeV Au+Au collisions.The solid curves in the right panel are the fit fuction of a short-range exponential and a wide gaussian terms.}
\label{fig:DeltaV2}
\end{figure}

Figure~\ref{fig:DeltaV2} left panel shows that $\Delta V_{2}\lbrace4\rbrace \approx \Delta\sigma_{2}'$ is consistent with zero within measurement uncertainty in 20-30\% central Au+Au collision at $\sqrt{s_{NN}}$ = 200 GeV. Hence, $\Delta V_{2}\lbrace2\rbrace=\Delta\delta_{2}$, which is shown in Figure~\ref{fig:DeltaV2} right panel. A function with a sharp expontential plus a wide Gausssian is used to describe the $\Delta V_{2}\lbrace2\rbrace$. The fitting result to $\Delta\delta_{2}$ is used to reconstruct the nonflow $\delta_{2}$.

\begin{figure}[htb]
\centering
\includegraphics[width=0.5\textwidth]{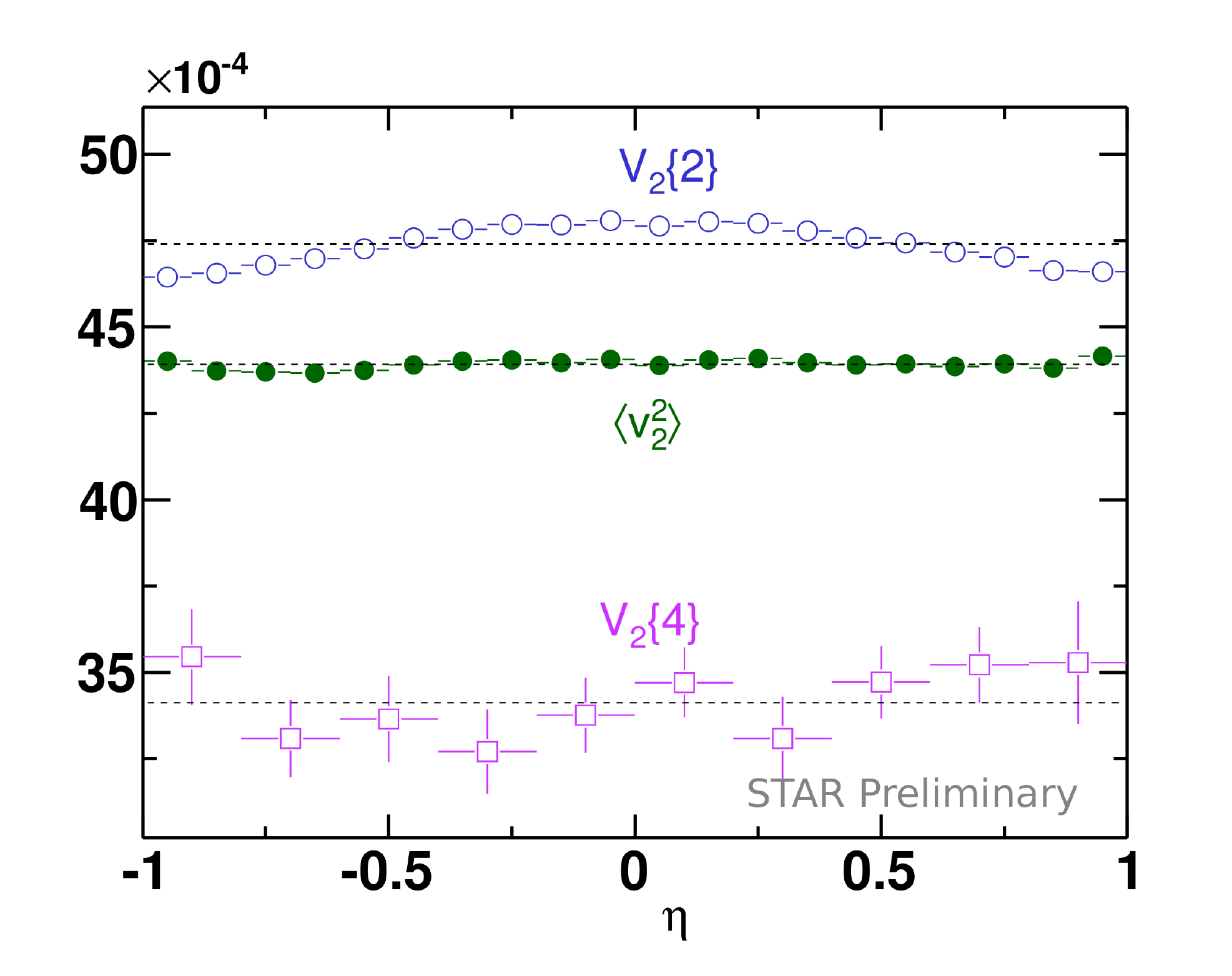}
\caption{The $\eta$ dependence of the raw two- and four-particle cumulants and the decomposed flow $\langle v_{2}^{2} \rangle$ of $v_{2}$ for 20-30\% centrality in Au+Au collisions at $\sqrt{s_{NN}}$ = 200 GeV.}
\label{fig:floweta}
\end{figure}

Subtracting the reconstructed nonflow from two-particle cumulant $V_{2}\lbrace2\rbrace$, the average flow and $\Delta\eta$-independent flow fluctuations are obtained. Figure~\ref{fig:floweta} presents the $\eta$ dependence of the raw two-particle cumulant $V_{2}\lbrace2\rbrace = \delta_{2} + \langle v_{2}^{2}\rangle$ as the blue open circles, the decomposed flow $\langle v_{2}^{2}\rangle$ as the green closed dots, and the square root of the raw four-particle cumulant $V_{2}\lbrace4\rbrace \approx \langle v_{2}^{2}\rangle - 2 \sigma_{2}^{2}$ as the purple open squares. The decomposed flow $\langle v_{2}^{2}\rangle$ appears to be independent on $\eta$. By taking the difference between $\langle v_{2}^{2}\rangle$ and $V_{2}\lbrace4\rbrace$, the $\Delta\eta$-independent flow fluctuation is estimated to be $\sigma_{2}/\langle v_{2} \rangle = 34$\%. Meanwhile, the difference between $V_{2}\lbrace2\rbrace$ and $\langle v_{2}^{2} \rangle$ is the nonflow. The nonflow is found to be $\delta_{2}/\lbrace v_{2}^{2} \rbrace = 5\%$ at $|\Delta\eta|>0.7$ for $p_{T}<2 $GeV/$c$ in 20-30\% central Au+Au collision at $\sqrt{s_{NN}} =$ 200 GeV.

\section{Summary}
\label{sec:sum}

The Beam Energy Scan results from 7.7 GeV to 39 GeV from STAR show that $v_{2}\lbrace4\rbrace$ increases with increasing collision energy. This is largely related to the weaker radial flow at the lower energy generating a smaller $\langle p_{T} \rangle$. In Au+Au 200 GeV collisions, the third harmonic flow $v_{3}$ shows strong $\Delta\eta$ dependence. The event-plane decorrelation and nonflow can contribute to the $\Delta\eta$ dependence. The isolation of flow and nonflow using 2- and 4-particle cumulants shows that the flow is $\eta$-independent. The relative nonflow $\delta_{2}/\langle v_{2}^{2} \rangle$ is found be to 5\%, and the flow fluctuation $\sigma_{2}/\langle v_{2} \rangle$ is 34\% for midrapdity $|\eta|<1$ and $0.15<p_{T}<2 $ GeV/$c$ in 20-30\% central Au+Au collision at $\sqrt{s_{NN}} =$ 200 GeV.





\end{document}